\RequirePackage{fix-cm}
\documentclass[smallextended]{svjour3}       
\smartqed  
\usepackage{graphicx}
\usepackage{lineno,hyperref}
\usepackage{amsmath, amssymb}
\usepackage{multirow}
\usepackage{graphicx, caption, color}
\usepackage{booktabs}
\usepackage{cite}
\begin{document}

\title{Preferred-frame Effects, the $H_0$ Tension, and Probes of Ho\v{r}ava-Lifshitz Gravity
}


\author{Nils A. Nilsson 
}


\institute{Nils A. Nilsson \at \\
              National Centre for Nuclear Research \\
              Pasteura 7 \\
              02-039, Warsaw, Poland\\
              \email{albin.nilsson@ncbj.gov.pl}
}

\date{Received: date / Accepted: date}

\maketitle

\begin{abstract}
We discuss implications on the $H_0$ tension due to preferred-frame effects in the context of Ho\v{r}ava-Lifshitz gravity. By using a combination of low-redshift data (Sne1a, elliptical and lenticular galaxies, GRB's, and quasars) we discuss the $H_0$ tension and its appearance as a preferred-frame effect, as well as present new constraints on the model parameter $\lambda$. Moreover, from the structure of the Friedmann equations, we argue that up to $36\%$ of the Hubble tension can be explained by Lorentz-violating effects in a Ho\v{r}ava-Lifshitz
    scenario.
\end{abstract}

\section{Introduction}
A long-standing problem in theoretical physics is the issue of quantum gravity, how to merge general relativity with quantum field theory. Although substantial effort has been put forth for several decades there is to date no clearly compelling candidate model. The main problem is that general relativity is not perturbatively renormalisable, which is a serious obstacle for standard quantisation techniques, leading to the breakdown of general relativity at small scales. 
    Many models have been proposed to deal with this problem, such as string theory and loop quantum gravity, and while these theories do resolve some of the problems of general relativity, there are few avenues available to test them~\cite{Quevedo:2016tbh,Girelli:2012ju}. Indeed, the fact that general relativity has passed every test so far indicates that it is an excellent model for the infrared (IR) behaviour or quantum gravity. This is natural since quantum-gravity effects are expected to
    emerge at energies close to the Planck energy. A natural course of action is then to study ultraviolet (UV) completion of general relativity, for example~\cite{ArkaniHamed:1998rs,ArkaniHamed:1998nn}. Another interesting proposal for a UV-complete theory of gravity is Ho\v{r}ava-Lifshitz gravity, which contains general relativity as an IR fixed point~\cite{Horava:2009uw}. The original formulation had problems such as ghost modes and instabilities, which were subsequently addressed in a
    series of papers, see for example~\cite{Blas:2010hb,Blas:2009qj,Blas:2009yd}. Since then, much work has been done on the subject, ranging from cosmological studies~\cite{Nilsson:2018knn,Mukohyama:2010xz,Kiritsis:2009sh,Frusciante:2015maa,Calcagni:2009ar,Appignani:2009dy,Cognola:2016gjy,Casalino:2018wnc}, dark energy~\cite{Saridakis:2009bv,Park:2009zr}, bouncing scenarios~\cite{Czuchry:2009hz,Brandenberger:2009yt}, and strong coupling~\cite{Charmousis:2009tc} among others. Ho\v{r}ava-Lifshitz gravity is a perturbatively
    renormalisable theory of gravity, which is accomplished by introducing a Lifshitz scaling between space and time in the UV~\cite{Horava:2009uw} which explicitly breaks Lorentz invariance. It is important to mention that Lorentz invariance is a building block of modern physics, and breaking it may seem counterinuitive. However, since the Planck scale and quantum gravity likely will contain completely new physics on quantum scales it is useful to not a priori \emph{assume} Lorentz
    invariance, which is a continuous symmetry, in this sector. 

    Recently, various measurements of the Hubble constant, $H_0$, have revealed a discrepancy between the value at high and low redshift, respectively. In fact, this discrepancy has been confirmed by many independent observations (using $\Lambda$CDM as a background model) at low (quasars~\cite{Birrer:2018vtm}, gravitational waves~\cite{Guidorzi:2017ogy,Feeney:2018mkj,Chang:2019xcb}, Cepheid stars~\cite{Riess:2018byc,Riess:2009pu,Zhang:2017aqn}) and high (Cosmic Microwave
    Background~\cite{Planck2018}, Baryon Acoustic Oscillations~\cite{Ross:2014qpa,Aubourg:2014yra}, the inverse distance ladder~\cite{Macaulay:2018fxi,Aylor:2018drw}) redshift. The difference in the value of the $H_0$ from these different observations lie around $4\%$ - $9\%$. Many scenarios have been put forth as explanations or alleviations of the $H_0$ tension, for example dynamical dark energy~\cite{Pan:2019gop}, screened fifth
    forces~\cite{Desmond:2019ygn}, the late decay of dark matter~\cite{Vattis:2019efj} and more, but the $H_0$ tension has proved diffcult to resolve. In this paper we investigate the presence of a preferred frame in the Universe and its effect of the $H_0$ tension. Working in a Ho\v{r}ava-Lifshitz model we constrain the discrepancy between our local frame and the preferred frame. Moreover, we suggest that part of the Hubble constant discrepancy is due to Lorentz violation in the ultraviolet
    regime.

\section{Ho\v{r}ava-Lifshitz Gravity}
Ho\v{r}ava-Lifshitz gravity is a proposal for a nonrelativistic theory of gravity, which breaks Lorentz invariance in the UV regime by introducing an anisotropic Lifshitz scaling between space and time of the form $t\rightarrow b^{-z}t$, $x^i\rightarrow b^{-1}x^i$ (breaking Lorentz invariance), where $z$ is a critical exponent~\cite{Horava:2009uw}. Lorentz invariance is restored for $z=1$, but in order to obtain power-counting renormalisability it is necessary to have $z\geq 3$ (for 3 spatial dimensions)~\cite{Wang:2017brl},
and we will set $z=3$. The theory is power-counting renormalisable and is a candidate theory of quantum gravity. In the IR, the theory reduces to that of general relativity. Much work has been done on this theory, including some early contributions to cure some of the original
inconsistencies~\cite{Pospelov:2010mp,Park:2009zr,Oost:2018tcv,Lu:2009em,Kiritsis:2009sh,Horava:2010zj,Frusciante:2015maa,Gumrukcuoglu:2017ijh,Dutta:2009jn,Czuchry:2009hz,Colombo:2015yha,Charmousis:2009tc,Blas:2011zd,Blas:2009qj,Blas:2009yd,Appignani:2009dy,Nilsson:2018knn}. The presence of the anisotropic scaling in the theory leads to a natural description using the Arnowit-Deser-Misner (ADM) formulation, in which the metric reads:
\begin{equation}
    ds^2 = -N^2dt^2+g_{ij}\left(dx^i+N^idt\right)\left(dx^j+N^jdt\right), 
\end{equation}
where $N$ and $N^j$ are the lapse function and the shift vector, which determine the foliation of spacetime by constant-time spacelike hypersurfaces. The breaking of Lorentz invariance in ultraviolet Ho\v{r}ava-Lifshitz gravity manifests as the appearence of a preferred foliation of spacetime, and the symmetry is most commonly assumed to be broken down to $t\rightarrow \xi_0(t)$, $x^i\rightarrow\xi^i(t,x^k)$. Then, the theory is endowed with the foliation-preserving diffeomorphism group,
denoted $\text{Diff}[M,\mathcal{F}]$, where $M$ is the manifold and $\mathcal{F}$ is the preferred frame. Given this, we can write down the most general form of the theory as:
\begin{equation}\label{eq:action1}
    S=\int d^3x dt N \sqrt{g}\left[K^{ij}K_{ij}-\lambda K^2-\mathcal{V}(g_{ij})\right],
\end{equation}
where $g$ is the determinant of the spatial metric, $\lambda$ is a running coupling and $\mathcal{V}$ is a potential. $K_{ij}$ represents the extrinsic curvature of the foliation. The potential term contains only dimension 4 and 6 operators which can be constructed from the spatial metric $g_{ij}$.
Under the so-called detailed balance and projectability conditions, the action reads~\cite{Dutta:2009jn}:
\begin{align}\label{eq:HLaction}
    \nonumber S &= \int dtd^3x\sqrt{g}N\Big[\frac{2}{\kappa^2}\left(K_{ij}K^{ij}-\lambda K^2\right)+\frac{\kappa^2}{2w^4}C_{ij}C^{ij}-\frac{\kappa^2\mu}{2w^2}\frac{\epsilon^{ijk}}{\sqrt{g}}\mathcal{R}_{il}\nabla_j\mathcal{R}^i_k +\\&+\frac{\kappa^2\mu^2}{8}\mathcal{R}_{ij}\mathcal{R}^{ij}+\frac{\kappa^2\mu^2}{8(1-3\lambda)}\Big(\frac{1-4\lambda}{4}\mathcal{R}^2+\Lambda \mathcal{R}-3\Lambda^2\Big)\Big]
\end{align}
where $\nabla_j$ is the spatial covariant derivative, $\epsilon$ is the totally antisymmetric tensor, and $\mu, w$, and $\kappa$ are dimensionful constants (mass dimension 1, 0, -1, respectively). Any higher-order terms are assumed to be Planck suppressed by $M_{\rm Pl}^{-n}$ (at order $n$), where $M_{\rm Pl}$ is the Planck mass. $C^{ij}$ is the Cotton tensor, and $\mathcal{R}_{ij}$ is the Ricci tensor related to the spatial metric. This action has been obtained from
(\ref{eq:action1}) by analytic continuation of the parameters $\mu$ and $\omega^2$, which enables positive values of the bare cosmological constant $\Lambda$, which does not occur in the original formulation of Ho\v{r}ava-Lifshitz gravity. 

Although the detailed-balance condition leads to a succinct action, there is an ongoing debate in the literature whether this formulation is too restrictive. In fact, there are a number of problems with the detailed-balance scenario, such as instabilities, strong coupling at low energies, as well as problems with the value of the cosmological constant~\cite{Kiritsis:2009sh, Colombo:2015yha, Nilsson:2018knn, Wang:2017brl}. As such, we choose to focus our efforts on the so-called
beyond detailed balance scenario~\cite{Charmousis:2009tc, Sotiriou:2009bx, 0264-9381-27-7-075005,Dutta:2009jn,Dutta:2010jh}, where it is possible to include more terms in the potential $\mathcal{V}$. Then, using the FLRW line element and populating the Universe with the canonical matter fields, the first Friedmann equation can be written:
\begin{align}\label{eq:friedmann}
    \left(\frac{\dot{a}}{a}\right)^2 = \frac{2\sigma_0}{3\lambda-1}(\rho_m+\rho_r)+\frac{2}{3\lambda-1}\left[\frac{\Lambda}{2}+\frac{\sigma_3K^2}{6a^4}+\frac{\sigma_4K}{6a^6}\right]+\frac{\sigma_2K}{3a^2(3\lambda-1)}.
\end{align}
Here, the objects $\sigma_i$ are arbitrary constants.


\section{Bounds on Ho\v{r}ava-Lifshitz Gravity from the $H_0$ tension}

\subsection{$H_0$ tension as a preferred-frame effect}
In~\cite{Chang:2019esw} the authors suggest that the discrepancy \cite{Riess:2018byc, Riess_2016} between the value of the Hubble parameter $H_0$ from CMB measurements and from local data is in fact a reference-frame artefact. Since Ho\v{r}ava-Lifshitz gravity is based on a preferred frame is it natural to also pose this question in this model. Following~\cite{Chang:2019esw} we use a flat FLRW metric and define a geodesic observer in the CMB frame as $v^\mu =
(\sqrt{1+(\zeta/a)^2},0,0,\zeta/a^2)$, where $\zeta$ is a parameter and $a$ is the FLRW scale factor. For this observer, the metric takes the form:
\begin{equation}
    ds^2 = -dT^2+a^2(T;Z)\left(dX^2+dY^2+(1+(\zeta/a)^2)/(1+\zeta^2)dZ^2\right). 
\end{equation}
Following~\cite{Chang:2019esw} we use the transformation which relates the Hubble constant in the local geodesic frame to that in the CMB frame:
\begin{equation}\label{eq:frametrans}
    \frac{H_0^{\rm CMB}}{H_0^{\rm local}} = \frac{1}{\sqrt{1+\zeta^2}}. 
\end{equation}
Hence, the local measurement has to be larger than or equal to its CMB counterpart. The two values will coincide when $\zeta\rightarrow 0$.
We find the low and high redshift values of the Hubble parameter using a Markov-Chain Monte Carlo analysis. Here, we adopt a methodology similar to~\cite{Lusso:2019akb} by using several different data sets from a wide, yet local, redshift range. For the local value of the Hubble constant we use the PANTHEON dataset of supernovae type Ia~\cite{Scolnic:2017caz}, along with expansion rates of elliptical and lenticular galaxies~\cite{Moresco15}, gamma-ray bursts~\cite{Liu:2014vda} and quasars~\cite{Risaliti:2015zla}. These sources are all within redshift range $0.01<z<8.2$, a large redshift range with multiple sources which we define as our ``local`` frame, as compared to the $z\sim 1040$ for the CMB frame. For details of the method, see~\cite{Nilsson:2018knn}. We find that $H_0^{\rm local} = 70.212 \pm 0.018$ km s$^{-1}$ Mpc$^{-1}$ at $99.7$\%. 
Moverover, for the high-redshift (early Universe) value of the Hubble parameter we use Planck CMB data~\cite{Planck2018}. We find that, at $99.7\%$, the Hubble constant is $67.23^{+5.0}_{-4.5}$ km s$^{-1}$ Mpc$^{-1}$, and using these two values of the Hubble constant in (\ref{eq:frametrans}) we find that the parameter $\zeta$, quantifying the discrepancy between the local frame and CMB frame, is (disregarding any negative values in order to keep $\zeta$ real):
\begin{equation}\label{eq:zetabound}
    0\leq \zeta^2 \leq 0.25.
\end{equation}

As suggested in~\cite{Chang:2019esw} we have found bounds on the parameter $\zeta$ from observations of the Hubble parameter. Thus, $\zeta$ defines a geodesic reference frame where the observed $H_0$ tension would emerge naturally.

\subsection{The $H_0$ tension and the Ho\v{r}ava parameter $\lambda$}
It is known that in Lorentz-violating field theories, the gravitational constant measured locally, $G^{\rm local}$ does not coincide with the cosmological one~\cite{Carroll:2004ai}. In fact, we will show that also the gravitational constant can be thought of as frame dependent, and we will give it a superscript, $G^{\rm CMB}$, to show that this is the value in the CMB frame. We may derive from (\ref{eq:friedmann}) that the value of the gravitational constant at different energy scales are
related by a single Ho\v{r}ava parameter~\cite{Dutta:2010jh}:
\begin{equation}
    G^{\rm CMB} = \frac{2}{3\lambda^{\rm CMB}-1}G^{\rm local},
\end{equation}
where the superscript on $\lambda$ is to highlight that it is the value of $\lambda$ at the time of recombination. The infrared fixed point $\lambda\rightarrow 1$ represents General Relativity, which is also when $G^{\rm CMB} = G^{\rm local}$. Clearly, in this scenario, dynamics will be different on cosmologcal scales. This also has implications for the Hubble tension. 
We can write down a general form of the first Friedmann equation in the two frames as:
    \begin{align}
        (H_0^{\rm CMB})^2 &= \frac{8\pi}{3}G^{\rm CMB} \rho_0\\
        (H_0^{\rm local})^2 &= \frac{8\pi}{3}G^{\rm local} \rho_0
    \end{align}
    where $\rho_0$ is the total energy density, which is the same in the two frames. On this basis we arrive to the same as (8) by dividing (9) by (10):
\begin{equation}\label{eq:H0_lambda}
    \left(\frac{H_0^{\rm CMB}}{H_0^{\rm local}}\right)^2 = \frac{G^{\rm CMB}}{G^{\rm local}} = \frac{2}{3\lambda^{\rm CMB}-1}.
\end{equation}
In the above relation we have to assume that Lorentz violation only \emph{contributes} to the Hubble tension rather than being the only cause of it. In light of this it would be more accurate to write the right-hand side as $2/(3\lambda^{\rm CMB}-1) + f(\mathbf{\theta})$, where $f(\theta)$ is an unknown function of one or more parameters.
We can now use available Hubble constant data to put constraints on the parameter $\lambda$, and also estimate the contribution of Lorentz violation to the Hubble tension.

\subsubsection{Constraints on $\lambda^{\rm CMB}$}
Currently, the most accurate measurements of the Hubble constant come from the local distance ladder ($74.03 \pm 1.42$ km s$^{-1}$ Mpc$^{-1}$~\cite{Riess:2018byc,Riess_2016,Riess:2019cxk}) and Planck CMB ($67.4 \pm 0.5$ km s$^{-1}$ Mpc$^{-1}$~\cite{Planck2018}). Ignoring any model dependence of these bounds we use Eq.~(\ref{eq:H0_lambda}) to find that $\lambda^{\rm CMB} = (0.86, 0.92)$ at $99.7\%$. Note that loverlooking the model dependence of these constraints is a strong assumption
(especially for the CMB value).
This can be compared to the limits on $H_0$ which we obtained in Ho\v{r}ava-Lifshitz using the beyond detailed balance formulation ($H_0^{\rm local} = 70.2\pm 0.02$ km s$^{-1}$ Mpc$^{-1}$, $H_0^{\rm CMB} = 67.2^{+5.0}_{-4.5}$ km s$^{-1}$ Mpc$^{-1}$). Indeed, using those values of the Hubble parameters we arrive at $0.95 \leq \lambda^{\rm CMB} \leq 1.16$ at $99.7\%$ confidence level.
The bounds on $\lambda^{\rm CMB}$ from local distance ladder and Planck data are problematic, since $1/3 < \lambda < 1$ generally leads to ghost instabilities in the IR limit~\cite{0264-9381-27-7-075005}, whereas the limit from the Hubble parameters found from MCMC analysis of Ho\v{r}ava-Lifshitz still overlap with a non-pathological region. 

From the same MCMC analysis which provided the bounds on the Hubble parameters we also obtained direct constraints on $\lambda^{\rm CMB} = 1.056\pm0.02$. This is encouraging, since the whole range lies in the non-pathological region for $\lambda$. A summary of all derived limits can be seen in Table 1.

\subsubsection{Constraints on the Hubble parameter}
Using available constraints on $\lambda$ we can get a value of the Hubble tension through Eq.~(\ref{eq:H0_lambda}). To our knowledge there is only one bound in the published literature, namely $\lambda = (0.97, 1.01)$~\cite{Dutta:2010jh}. Using this we find that $H^{\rm CMB}/H^{\rm local} = (0.98, 1.01)$. This can be compared to the value from local distance ladder and Planck CMB measurements, where the same ratio works out to $H^{\rm CMB}/H^{\rm local} = (0.89, 0.94)$. The
central value of this interval is $0.915$, leading to a Hubble tension of $8.5\%$. Taking a conservative approach we use the upper bound of the calculated Hubble ratio from~\cite{Dutta:2010jh} and comparing to the observed $8.5\%$ Hubble tension means that in this scenario, Lorentz violation can be the source of up to $12\%$ of the Hubble tension. 
It is important to keep in mind that the constraints on $\lambda$ in \cite{Dutta:2010jh} were derived using a large set of cosmological data from both high and low redshift, and the resulting value must be considered an \emph{average} $\lambda$. However, since it is the only (to our knowledge) published constraint on $\lambda$ we have used it, keeping in mind the above discussion. Since $\lambda$ runs with energy we can assume that it was larger in the early Universe and therefore likely
contributes more to the observed Hubble tension than our bound of $\leq 12\%$ indicates.

We may also use our derived constraints on $\lambda^{\rm CMB} = (0.95,1.16)$ and assuming Lorentz violation is the only source of the Hubble tension, the corresponding tension is $3.8\%$. By again comparing to the observed $8.5\%$ this we can infer that, at $99.7\%$ confidence level, Lorentz violation can be the source of up to $44.7\%$ of the Hubble tension. 

Finally, we may also use our direct constraint $\lambda^{\rm CMB} = 1.056\pm0.02$. In order to find the most conservative estimate we use the upper bound of $\lambda^{\rm CMB}$, which combined with the  measured Hubble tension of $8.5\%$ leads to a possible contribution of Lorentz violation of up to $38\%$. This is our
main result.

\begin{table}[h!]
\begin{center}
\caption{Summary table of constraints on preferred-frame effects on the Hubble tension, as well as constraints on $\lambda$ and the Lorentz violation contribution to the Hubble tension.}
    \label{tab:results}
    \begin{tabular}{ll}
        \textbf{Preferred frame} & \textbf{Constraint}\\
        \toprule%
        Ho\v{r}ava model + Planck CMB~\cite{Planck2018} & $0\leq \zeta^2 \leq 0.25$ \\
        \textbf{Constraints on $\lambda^{\rm CMB}$ from $H_0^{\rm CMB}$} & \textbf{Constraint}\\
        \midrule%
        Ho\v{r}ava model + Planck CMB~\cite{Planck2018} & $\lambda^{\rm CMB} = 1.056 \pm 0.02$ \\
        Derived from Ho\v{r}ava bounds on $H_0^{\rm CMB}$ & $0.95 \leq \lambda^{\rm CMB} \leq 1.16$\\
        \textbf{Hubble tension data} & \textbf{Lorentz violation contribution}\\
        \midrule%
        $\lambda$ from~\cite{Dutta:2010jh} + MCMC analysis & $\leq12\%$\\
        Derived from $H_0^{\rm CMB}$ + MCMC analysis & $\leq44.7\%$ \\
        Ho\v{r}ava model + Planck CMB~\cite{Planck2018} & $\leq38\%$.
    \end{tabular}    
\end{center}
\end{table}

\section{Discussion \& Conclusions}
In this article we have provided new bounds on preferred-frame effects and Ho\v{r}ava-Lifshitz gravity through the $H_0$ tension. Using a value for $H_0$ in the CMB frame for Ho\v{r}ava-Lifshitz gravity along with a local value, both from our own Markov-Chain Monte Carlo analysis, we were able to place bounds on $\zeta$, which determines the transformation from the CMB frame to the geodesic frame completely.
In~\cite{COLEMAN1997249} the authors point out an interesting consequence of a preferred frame. Indeed, if the frame $\mathcal{F}$ moves relativistically with respec to the CMB frame, there would be an observable effect in the form of a dipole anisotropy of high-energy cosmic rays in the sky. In fact, according to~\cite{Abbasi_2014}, there are indications of this at intermediate scales at $3.4\sigma$ significance, with no known specific sources in the
direction of the hotspot. These results are based on the observation of the northern hemisphere between May 2008 to May 2013, yielding 72 cosmic-ray events with energies higher than $57$ EeV.

Moreover, we have also founds new bounds on the Ho\v{r}ava-Lifshitz parameter $\lambda$ using Hubble constant data and our own MCMC simulations using cosmological data. We find that some of these bounds overlap significantly with regions of $\lambda$ known to lead to ghost instabilities in the infrared limit of the theory, but that some bounds also cover a non-pathological parameter space.
Moreover, we have used available bounds on $\lambda$ to estimate how much Lorentz-violating effects could contribute to the Hubble tension. Most significantly, we find that when using our own bounds on $\lambda$ from the beyond detailed balance scenario along with a MCMC method and Planck CMB data, Lorentz violation can contribute to up to $38\%$ of the Hubble tension. 
Therefore it would make sense to also consider Lorentz-violating field theories in the search to find an explanation for
the Hubble tension.

\section*{Acknowledgements}
NAN is grateful to Mariusz P. D\c{a}browski and Viktor Svensson for useful discussions. NAN was partly funded by the NCBJ Young Scientist Grant MNiSW 212737/E-78/M/2018.

%
\section*{Conflict of interest}
The authors declare that they have no conflict of interest.

\bibliographystyle{spphys}       
\bibliography{references_HL_cosmicrays}   
%
%

\end{document}